\documentclass[groupedaddress, superscriptaddress, twocolumn, pra]{revtex4-1}
\usepackage{graphicx,amsfonts,mathrsfs,amsmath}
\usepackage{physics}
\usepackage[usenames,dvipsnames]{color}
\usepackage[colorlinks=true,linkcolor=BrickRed,citecolor=BrickRed,urlcolor=Blue]{hyperref}

\renewcommand\phi\varphi

\begin{document}

\title{Improved entanglement detection with subspace witnesses}

\author{Won Kyu Calvin Sun} 
\email{wksun@mit.edu}
\affiliation{Research Lab of Electronics and Department of Nuclear Science and Engineering, Massachusetts Institute of Technology, Cambridge, MA 02139, USA}
\author{Alexandre Cooper}
\affiliation{Institute for Quantum Computing, University of Waterloo, Waterloo, ON, Canada, N2L 3G1}
\author{Paola Cappellaro}
\affiliation{Research Lab of Electronics and Department of Nuclear Science and Engineering, Massachusetts Institute of Technology, Cambridge, MA 02139, USA}

\begin{abstract}
%Entanglement, while being critical in many quantum applications, is difficult to characterize  experimentally. While entanglement witnesses based on the fidelity  to the target entangled state  are efficient detectors of entanglement, they in general underestimate the amount of entanglement due to  errors during state preparation and measurement. Here, to improve  entanglement detection, we introduce a `subspace' witness that can detect a broader class of entangled states  than the conventional state-fidelity witnesses, while still remaining more efficient than state tomography. We experimentally demonstrate the advantages of the subspace witness by generating and detecting entanglement with a hybrid, two-qubit system composed of electronic spins in diamond. We further extend the notion of  subspace witness  to specific genuine multipartite entangled (GME) states such as GHZ, W, and Dicke states, and motivate the choice of the metric based on quantum information tasks such as entanglement-enhanced sensing. We expect the straightforward and efficient implementation of subspace witnesses would be beneficial in detecting specific GME states  in noisy, intermediate-scale quantum processors with a hundred qubits.

%2019-11-14 wksun (after PRA review) : slightly modified abstract
Entanglement, while being critical in many quantum applications, is difficult to characterize experimentally. While entanglement witnesses based on the fidelity to the target entangled state are efficient detectors of entanglement, they in general underestimate the amount of entanglement due to local unitary errors during state preparation and measurement (SPAM). Therefore, to detect entanglement more robustly in the presence of such control errors, we introduce a `subspace' witness that detects a broader class of entangled states with strictly larger violation than the conventional state-fidelity witness at the cost of additional measurements while remaining more efficient with respect to state tomography. We experimentally demonstrate the advantages of the subspace witness by generating and detecting entanglement with a hybrid, two-qubit system composed of electronic spins in diamond. We further extend the notion of subspace witness to specific genuine multipartite entangled (GME) states detected by the state witness, such as GHZ, W, and Dicke states, and motivate the choice of the metric based on quantum information tasks, such as entanglement-enhanced sensing. In addition, as the subspace witness identifies the many-body coherences of the target entangled state, it facilitates (beyond detection) lower bound quantification of entanglement via generalized concurrences. We expect the straightforward and efficient implementation of subspace witnesses would be beneficial in detecting specific GME states in noisy, intermediate-scale quantum processors with a hundred qubits.
\end{abstract}

\maketitle

\section{Introduction}

Entanglement describes quantum correlations with no classical analog, and it underpins many advantages of quantum devices over classical computation, communication and sensing~\cite{Bennett93, Bennett96, Bollinger96}, while also being central in many physical phenomena such as phase transitions~\cite{Amico08}.  However, entanglement  is difficult to characterize both theoretically and even more so experimentally. 
The most direct way requires performing quantum state tomography (QST)~\cite{Amiet99, DAriano03} to obtain the density operator $\rho$  describing the state, and then use one of several metrics of entanglement that have been proposed. Unfortunately, QST requires a number of measurements that scales exponentially with the qubit number $n$, such that for large $n$ QST becomes intractable.  Even for small systems, errors in  state preparation and measurement (SPAM) compound the difficulty in identifying $\rho$ with high accuracy~\cite{Steane03, Merkel13, Bogdanov16, Bantysh19a}.

When the goal is more simply to detect whether entanglement is present or not, an attractive alternative is to use so-called entanglement witnesses $W$~\cite{Peres96, Horodecki96}. In contrast to QST, the resources needed to measure an entanglement witness typically scale more favorably with respect to $n$. The witness operator $W$ is designed to `witness', that is, detect a specific entangled state $|\psi\rangle$:  its expectation value $\langle W_\psi \rangle = \text{tr} (\rho W_\psi) $ is negative for some entangled states, while it is positive for all separable states. 
While designing a witness for an arbitrary entangled state is difficult, since this would solve the separability problem~\cite{Guhne09, Gurvits03a, Ioannou06, Gharibian08}, for NPT entangled states $|\psi\rangle$ (states that have negative eigenvalues under positive partial transpose)~\cite{Bergmann13}, such as the well-known Bell, GHZ, W, and Dicke states, the witness is based on state fidelity, $\langle F_\psi\rangle\equiv \bra\psi\rho\ket\psi$:
\begin{equation} \label{eq:stateWitness}
\begin{split}
W_\psi & = \alpha \openone  -  |\psi\rangle \langle \psi|.\\
%\langle W_\psi(\phi) \rangle & = \alpha  - \langle F(\phi) \rangle \\
%F(\phi) & \equiv |\psi(\phi)\rangle \langle \psi(\phi)|\\
\end{split}
\end{equation}
Here $\alpha$ is the squared maximum overlap of $|\psi\rangle$ with all possible separable states~\cite{Bourennane04} so as to conservatively detect entanglement.
%2019-11-14 wksun (after PRA review): new sentence including suggested citations from First Referee
This improved experimental feasibility---requiring only the measurement of state fidelity---has led to successful demonstrations of entanglement detection across many platforms~\cite{Barbieri03, Bourennane04, Friis18, Wei19, Omran19, Song19, Bradley19} as well as theoretical improvements and modification of the witness to further improve experimental feasibility~\cite{Sciara19}.

However, one immediate drawback of such `state' witnesses $W_\psi$ is that they can severely underestimate the amount of entanglement actually present in $\rho$. While errors in state preparation  can indeed lower the entanglement from the target state $\rho=|\psi \rangle \langle \psi |$, some common errors such as local and unitary errors will not change the amount of entanglement.  However, the typical witnesses $W_\psi$ of the form in Eq.~\ref{eq:stateWitness} will not capture this entanglement.

Here we use a solid-state 2-qubit system to investigate the advantages and limits of entanglement witnesses in the presence of SPAM errors. To achieve more robust entanglement detection, we introduce a new metric, which we call `subspace' witness $\langle W_s \rangle$, that can capture a larger share of entangled states generated in the presence of unitary, local errors. We compare state $\langle W_\psi\rangle$ and subspace $\langle W_s \rangle$ witnesses,  observing  an improvement in the detection of entanglement by $\langle W_s \rangle$, that can even provide a stricter bound on entanglement quantification. 
As an extension, we discuss the subspace witness measurement for specific genuine multipartite entangled (GME) states, such as GHZ, W, and Dicke states,  that are compatible with state-fidelity witnesses. 
We find that quite broadly the subspace witness allows identifying all the  many-body coherences, which are often of interest in practical applications such as  quantum sensing.

\begin{figure}
  	\includegraphics[width=1\linewidth]{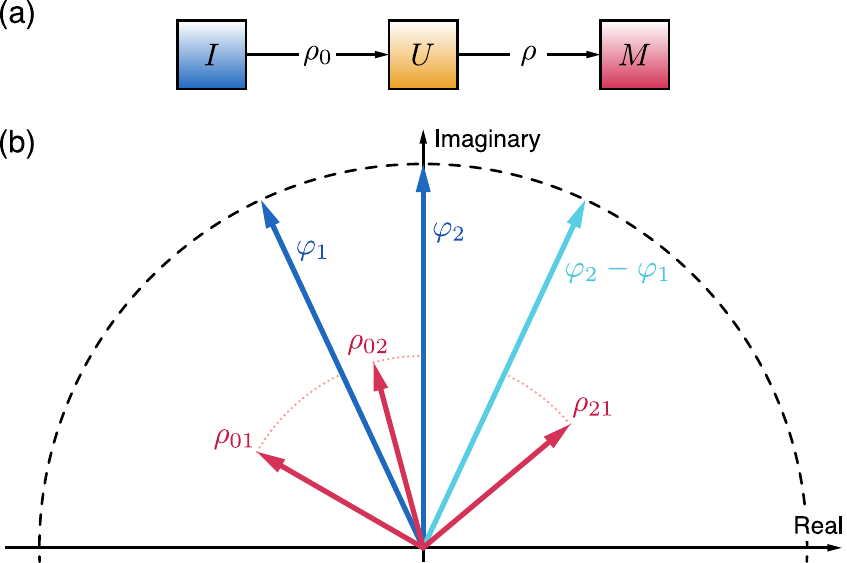}
 	\caption{\textbf{Entanglement detection} 
	(a) General entanglement detection scheme: to initialize state to $\rho_0$ (\textit{I}), entangle  $\rho_0\mapsto\rho$ (\textit{U}), and measure $\rho$ (\textit{M}) at desired settings parameterized by operator $M(\vec{\phi})$, where typically the fidelity measurement $M(\vec{\phi})=F_\psi(\vec{\phi})=|\psi(\vec{\phi})\rangle \langle \psi(\vec{\phi})|$ is desired.	
	(b) Visualizing (improved) fidelity-based entanglement detection: as seen from equations~\ref{eq:BellCoherence} and~\ref{eq:explicitSubspaceWitness}, fidelity can be improved by increasing the overlap between vectors describing the many-body coherences $\rho_{jk}$ and vectors describing the measurement or equivalently the target entangled state $|\psi(\vec{\phi})\rangle$. Given that $\rho$ is a priori unknown, one knows not the optimal fidelity measurement operator $F_\psi(\vec{\phi})$; furthermore, given that in general $\rho$ is mixed, a single fidelity measurement---and thus a typical `state' witness measurement $\langle W_\psi\rangle$ (Eq.~\ref{eq:stateWitness})---cannot reveal the true (i.e., maximum) coherences $|\rho_{jk}|$. Thus we discuss a `subspace' witness measurement $\langle W_s\rangle$ (Eq.~\ref{eq:subspaceWitness}), relying on multiple fidelity measurements, for improved entanglement detection by identification of true coherences $\rho_{jk}$. Here the schematic illustrates a single fidelity measurement $F_\psi(\phi_1,\phi_2)$ with respect to a 3-qubit W-state as the target $|\psi\rangle=(|0\rangle+e^{-i\phi_1}|1\rangle+e^{-i\phi_2}|2\rangle)/\sqrt{3}$---which fixes the blue, cyan arrows---given the state $\rho$ whose (unknown) coherences $\rho_{jk}$ are shown in pink.
	%Here the schematic illustrates, given a 3-qubit W-state as the target, a single fidelity measurement $F_\psi(\phi_1,\phi_2)$ (which fixes the blue and cyan arrows) to overlap with the a priori unknown coherences of interest $\rho_{jk}$ (pink arrows) where $\ket{j}$ denotes the state with $(j+1)$-th qubit excited.
	%\textbf{the following sentence is too long, convoluted and needlessly complicated. Make sure you clarify what you want to convey in your mind and then try to rewrite.}Schematic of many-body coherences ($\{ \rho_{jk} \} $) as extracted by fidelity-based entanglement witnesses, e.g., in figure probing tripartite (entangled) state $\rho$ (subspace dimension $d=3$) with fidelity to $|\psi\rangle = |W(\phi_1,\phi_2)\rangle$: since in general---due to $\rho$ unknown---such witnesses underestimate $\rho_{jk}$'s, a `subspace' witness could be used to extract the maximum $|\rho_{jk}|$'s, at the cost of measurements increasing as $d(d-1)+1$, thereby remaining efficient with respect to state tomography.
	}
	\label{fig:Figure1SubspaceWitnessConcept}
\end{figure} 

\section{Witnessing two-qubit entanglement}

\subsection{State witnesses $W_\psi$}

For a two qubit system, there are four canonical maximally entangled states, the Bell states~\cite{Bell64, Clauser69, Bennett96}. The Bell states $\{ |\Phi^\pm \rangle, |\Psi^\pm \rangle \}$  form an orthogonal basis, thus  any  state (and in particular entangled states) can be written in terms of their superpositions. 
Choosing the computational basis to describe the energy eigenbasis, we can explicitly write a Bell state as $\ket{\Phi^\pm(\Psi^\pm)} = (\ket{k}\pm \ket{\bar{k} })/\sqrt2$, with $\ket{k} = \ket{00}(\ket{01})$  and $\ket{\bar{k}}$ the corresponding spin-flipped states. Each pair of Bell states, $\ket{\Phi^\pm}$ and $\ket{\Psi^\pm}$, span a subspace with constant energy. For many applications, such as entanglement-enhanced quantum sensing~\cite{Bollinger96, Meyer01, Cooper18} or decoherence-protected subspaces~\cite{Palma96, Lidar98, Kwiat00, Fortunato02, Cappellaro06, Cappellaro07},  states inside this subspace are equally beneficial. In particular, we can identify the family of maximally entangled states inside such subspaces, parametrized by a phase $\phi$,
\begin{equation}
\ket{\Phi(\phi)}=\cos(\phi/2)\ket{\Phi^+}+i\sin(\phi/2)\ket{\Phi^-},
\label{eq:family}	
\end{equation}
and similarly for $\ket{\Psi^\pm}$. Here  $\phi$ describes the phase degree of freedom  that leaves unchanged  the state desired properties (e.g., for enhanced sensing or decoherence-protected memory respectively). 

Fixing $\phi$, we can build a canonical `state' witness as 
in Eq.~\ref{eq:stateWitness} (with $\alpha=1/2$). This is a good witness to detect the presence of two-qubit entanglement in any state $\rho$,  given that all two-qubit entangled states are NPT. The expectation value of the witness depends on the state fidelity, $W_\psi=1/2-F_\psi$, where the state fidelity is a function of $\phi$:
%Then we explicitly see the dependence of entanglement detection on $\phi$ in the state fidelity :
\begin{equation} \label{eq:BellFidelity}
\begin{split}
%\langle W_\psi(\phi) \rangle & = \alpha - \langle \psi (\phi) | \rho | \psi(\phi) \rangle \\
\langle F_\psi\rangle=\langle \psi (\phi) | \rho | \psi(\phi) \rangle  & = P + C(\phi) \leq  P + | \rho_{k\bar{k}} |.
\end{split}
\end{equation}
Here $P = 1/2 (\rho_{kk} + \rho_{\bar{k} \bar{k} })$ is the sum of populations in the $\ket{k},\ket{\bar k}$ subspace and
\begin{equation} \label{eq:BellCoherence}
\begin{split}
C(\phi)  & = \text{Re}(\rho_{k\bar{k}})\cos(\phi) + \text{Im}(\rho_{k\bar{k}})\sin(\phi)  \\
		& = |\rho_{k\bar{k}}|\cos(\phi + \theta_{k\bar{k}}) 
\end{split}
\end{equation}
are the related coherences, with $\tan(\theta_{k\bar{k}}) =  \text{Im}(\rho_{k\bar{k}})/\text{Re}(\rho_{k\bar{k}})$. 
The coherence $C(\phi)$ is maximum only for $\theta_{k\bar{k}}=-\phi$. Unfortunately, $\theta_{k\bar{k}}$ might be unknown due to the unitary component of SPAM errors. Then, while 
 $P<1/2$ always reflects a suboptimal (or absent) entanglement, $C(\phi)$ might even be negative although the state is maximally entangled. Not only this leads to an underestimate of the entanglement unless $\theta_{k\bar{k}}=-\phi$, but more critically, of the useful entanglement for many quantum tasks,  as often the exact value of  $\theta_{k\bar{k}}$ is  unimportant. 
%Thus, given that a priori the coherence $\rho_{k\bar{k}}$ is always unknown, one cannot choose the optimal $\phi$ in $W_\psi(\phi)$ to extract $\max_{\phi} C(\phi) = |\rho_{k\bar{k}}|$. 
%In other words, entanglement detection by the state witnesses $W_\psi$ in general will be underestimated due to unitary SPAM errors.

\subsection{Subspace witnesses $W_s$}

As a way to improve upon entanglement detection by state witnesses, we propose a `subspace' witness measurement that becomes insensitive to some unitary SPAM errors:
\begin{equation} \label{eq:subspaceWitness}
\begin{split}
\langle W_s \rangle 	& = \min_\phi \langle W_\psi \rangle =  \alpha - \max_\phi \langle \psi (\phi) | \rho | \psi(\phi) \rangle. \\
\end{split}
\end{equation}

We call this a `subspace' witness as for any state in the subspace spanned by the relevant entangled-state basis, the witness is able to detect whether it is entangled or not. 
%in the sense that it yields the fidelity maximized over the entangled subspace. 
The subspace witness can thus be considered an intermediate metric between state witnesses and entanglement measures: while entanglement measures provide a quantitative estimate of the entanglement amount, typically by optimizing over all possible local unitaries, the subspace witness optimizing  over a set of local unitaries that are of relevance for a particular quantum information tasks, thus detecting entanglement more robustly than the state witness, while still maintaining an efficient protocol.

Indeed, to experimentally obtain the subspace witness by maximizing the fidelity over the subspace, one needs to  simply perform multiple state witness (or fidelity) measurements.
That is, improved entanglement detection comes  at the cost of additional measurements. 
Still, the number of measurements is  much smaller than for QST. For two-qubit entanglement, Eq.~\ref{eq:BellFidelity} and ~\ref{eq:BellCoherence} show that there are $3$ unknowns: $\text{Re}(\rho_{k\bar{k}})$, $\text{Im}(\rho_{k\bar{k}})$, and $P$. Thus, three measurements, e.g., at $\phi=\{0, \pi/2, \pi\}$, fully identify $P$ and $|\rho_{k\bar{k}} |$, thus yielding the subspace-optimized entanglement witness. While the advantage for two qubits is not large, it quickly becomes substantial for larger systems, as we will see in Sec.~\ref{sec:GME}.

Having discussed the idea of subspace witness  we now describe our experimental system and the experimental protocol to measure $W_s$.

\section{Experimental Generation and Detection of Entanglement}

\subsection{Entanglement Generation}

\begin{figure*}
  	\includegraphics[width=1\linewidth]{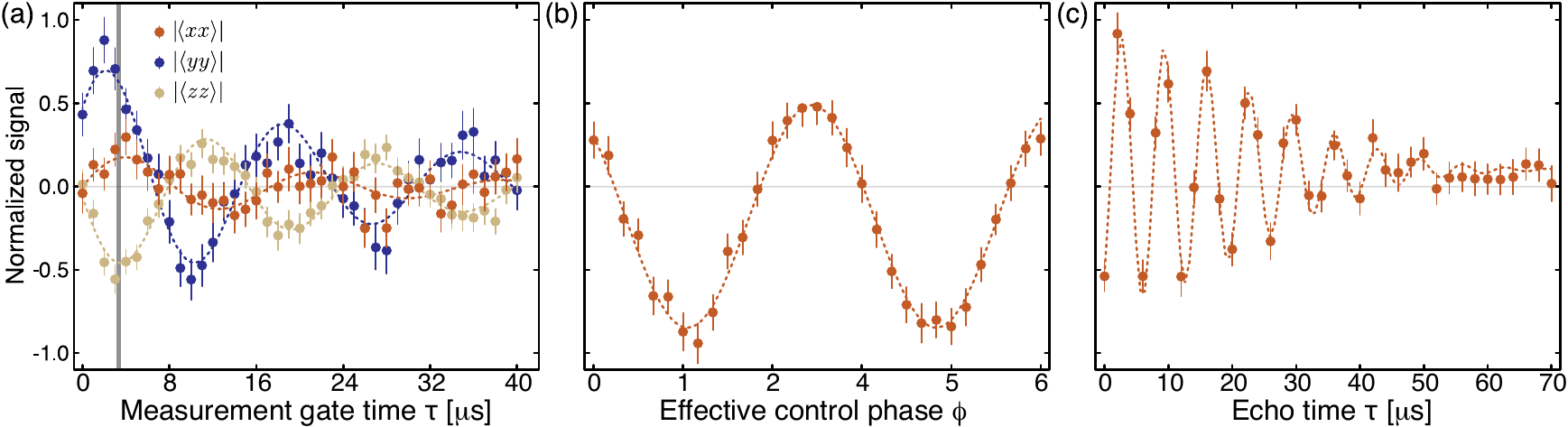}
 	\caption{\textbf{Demonstration of witness measurements $\langle W_\psi \rangle$ and $\langle W_s \rangle$ for $d=2$ target entangled state} (a) We measure the `state' entanglement witness measurement, based on Bell state fidelity, which successfully detects entanglement by $\langle W_\psi \rangle =  \alpha - \langle \Phi^+ |\rho| \Phi^+ \rangle = -0.07421(4)$. Grey vertical line denotes the optimal measurement gate time  that would yield the desired two-body correlators $\langle \sigma^j_1\sigma^j_2 \rangle $ in the absence of decoherence. To account for decoherence, the signal is fit (dotted lines) to exponentially decaying  oscillations  (Eq.~\ref{eq:measOpZZ}) with characteristic decay time $T=25\mu s$. Given the short optimal gate time, we see little difference when accounting or not for the decay. The measured two-body correlations are $|\langle \sigma^x_1\sigma^x_2 \rangle|=0.2142(1)$, $|\langle \sigma^y_1\sigma^y_2 \rangle|=0.5857(2)$, and $|\langle \sigma^z_1\sigma^z_2 \rangle|=0.4970(0)$. 
(b) Sweeping the control phase $\phi$ reveals oscillations between the real and imaginary part of the coherence $\rho_{14}$. By fitting the oscillations (dotted line) we extract the coherence amplitude and calculate the entangled state fidelity maximized over the Bell subspace, thereby improving entanglement detection by  $\langle W_s \rangle = \alpha - \langle \Phi|\rho| \Phi \rangle = -0.1827(4)$.
 (c) Measuring the spin echo after preparing the entangled state also yields the subspace witness, as the coherence $\rho_{14}$ time evolution is equivalent  to sweeping a phase $\phi \equiv \nu \tau$; this detection method  further estimates the time-scale of (detectable) entanglement. The two electronic spin system in diamond, after entangled state preparation to $\rho$, decohere under the spin echo pulse sequence, yielding a characteristic decay time $T_2=31(3)\mu s$ when fitted to a Gaussian decay (dotted line). As the population  $P=0.371$ is constant over the timescale of experiment, as measured independently, we witness entanglement until $\tau^*\leq T_2 \ln(C(0)/(\alpha-P))^{1/p}=33(3)\mu s$.}
%	the observed double-quantum coherence $|\langle00|\rho(\tau)|11\rangle|$---and not fidelity---detects entanglement up to $\tau^*_\text{LB} = T_2 \log(\rho_{jk}/\rho_{jk,\text{threshold}})^{1/p} = 21 \mu s$, while a fidelity threshold will yield a longer time until which entanglement is detected.}
	\label{fig:Figure2WitnessMeasurements}
\end{figure*}

We generated entanglement in a solid state two-qubit system, comprising two electronic spin impurities in diamond. The qubits are given by two electronic-spin levels of a single nitrogen-vacancy (NV) center, weakly interacting with a nearby, optically dark, electronic spin-1/2 defect in diamond. This spin-qubit has been earlier characterized as an electronic-nuclear spin defect~\cite{Cooper18a}. Here we neglect the nuclear spin degrees of freedom and remove the  electronic spin dependence on their state  by applying two-tone microwave pulses detuned by the nuclear hyperfine strength.

To experimentally prepare an entangled state $\rho$, we follow the protocol described below whose details are provided in~\cite{Hartmann62, Cooper18}. 
Starting from the thermal equilibrium state of the two spins, we first initialize the two-qubit system to $\rho_0$, via a Maxwell demon-type cooling scheme.
Specifically, we first initialize the NV spin to a high purity state using spin non-conserving optical transitions under laser illumination~\cite{Redman92, Gruber97}.
Then, we swap its state with the dark electronic spin X, 
and re-initialize the NV. 
After the initialization step $I$, we apply an entangling operation $U$ to prepare the desired state $\rho$ (Fig.~\ref{fig:Figure1SubspaceWitnessConcept}(a)).

Both the SWAP in $I$ and the entangling gate in $U$ are achieved by Hartmann-Hahn cross-polarization (HHCP), which exploits the spin coupling described by $\mathcal{H}_\text{int}= d \sigma_1^z \sigma_2^z /2$ as follows~\cite{Hartmann62}. After a global $\pi/2$-rotation to the transverse plane, we drive both spins with driving strength $\Omega_\text{NV}$ and $\Omega_\text{X,j}$, respectively (we use two driving fields $j=1,2$ for the dark spin to drive both nuclear hyperfine transitions). By tuning the driving strengths and frequencies ($\delta\omega_\text{NV}$ and  $\delta\omega_\text{X,j}$ for $j$-th nuclear spin state) the two spins can be brought on resonance in the dressed basis. This allows for polarization exchange thanks to the coupling $d$ between the two spins, despite the large energy mismatch in the lab frame. More generally, by a judicious choice of driving phases and timing, the HHCP scheme can realize two-qubit conditional gates~\cite{Belthangady13, London13, Knowles16, Rosenfeld18, Cooper18}.

In the case of ideal control, i.e., given no detuning in driving ($\delta\omega_\text{NV} = \delta\omega_\text{X,j} = 0$) and perfect Rabi matching ($\Omega_\text{NV}=\Omega_\text{X,j}$), the HHCP scheme  engineers  evolution under $\mathcal{H}_\text{HH} = d (\sigma^x_1\sigma^x_2 \pm \sigma^y_1\sigma^y_2)/4$. 
%todo: check sub/sup
Then a driving time of $dt=\pi/4$ would realize the gate $U = \sqrt{\text{iSWAP}}$ useful for generating entanglement. Similarly, with $dt=\pi/2$, ideal control would implement the $\text{iSWAP}$ gate as needed the initialization gate $I$. 

Due to intrinsic limits in the NV polarization process~\cite{Robledo11a}, and control errors and decoherence during the swapping operation, the state prepared by $I$ has sub-unit purity ($\text{tr}(\rho_0^2)<1$).
In addition, control in $I$ and $U$ is limited not only by decoherence, but also by local unitary rotations. 
Then, the prepared state $\rho$ might not be as desired, and a state witness might underestimate the entanglement present. To compound these issues, as we explain in the next section, similar control operations are needed to measure entanglement, given the available observables. Thus to partially relieve these SPAM errors, we show that it is beneficial to use subspace witnesses.

In the following, we  measure both the state $\langle W_\psi \rangle$ and subspace witnesses $\langle W_s \rangle$, and observe the presence of unitary, local errors, thus motivating the use of $\langle W_s \rangle$.

\subsection{Witness measurements}

Here we discuss the experimental protocol for measuring both the state witness $\langle W_\psi \rangle$ and subspace witness $\langle W_s \rangle$ in our system.

Ideally, to measure the Bell state fidelity $F_\psi$ that enters in $\langle W_\psi \rangle$ we would want to measure the observable $M_2'   = |\Phi^+\rangle \langle \Phi^+|$, such that the experimental signal would directly yield the Bell fidelity: $S=\text{tr}(M_2 U^\dag \rho U) = \langle \Phi^+| \rho |\Phi^+\rangle $. While unfortunately this is typically not possible, we can obtain $M_2'$ by a suitable unitary transformation of any 
joint projective measurement operator on the two qubits~\cite{Lloyd01}. For example,  consider an experimental system with the joint projective measurement operator $M_{n=2}=|0\rangle \langle 0 |^{\otimes n}$. 
Prior to measurement, we evolve the  state of interest $\rho$ under a unitary disentangling gate $U^\dag$, such as $U = \text{C}_1\text{NOT}_2 \cdot \text{H}_1$ (where $H_i$ is the Hadamard gate applied on qubit $i$).
This  is equivalent to transforming the bare measurement operator $M_2$ into $M_2' = U M_2 U^\dag = |\Phi^+\rangle \langle \Phi^+|$, as desired. 

Unfortunately, our hybrid system lacks a joint projective measurement operator such as $M_2$, as we can only directly measure the state of the NV center, $M=\ket0\!\bra0\otimes\openone$. In this case, even with universal control on the two-qubit system, we cannot measure the Bell fidelity in a single measurement. Therefore, here we introduce a protocol to reconstruct the Bell state fidelity that exploits measuring three correlators, $\langle \sigma_1^\alpha\sigma_2^\alpha\rangle$ with $\alpha=\{x, y, z\}$. 

In the experiments, we measure $M' = U M U^\dag =\sigma_1^z \sigma_2^z  $, with $U = \text{C}_1\text{NOT}_2$ and the  bare operator $M = \sigma_1^z = \ket0\!\bra0- \ket1\!\bra1$ obtained from the difference signal of measuring the NV states $0$ and $-1$. The CNOT gate is realized by exploiting evolution under the two-spin interaction Hamiltonian $\mathcal{H}_\text{int}$,  via the pulse sequence $R_x(\frac{\pi}{2}) \cdot e^{-iHt} \cdot R_y(\frac{\pi}{2})$ with $t=\pi/4d$ and $R_\alpha(\theta)$ collective rotations of the two spins along the axis $\alpha$ for an angle $\theta$. The  other two correlators $M'=\sigma_1^\alpha \sigma_2^\alpha$ for $\alpha=\{x,y\}$ can be measured by adding a global $R(\pi/2)$-rotation along the $\{y,x\}$ directions respectively. 

Given a joint projective measurement  as discussed above,  globally rotating the  disentangling gates along z, $M_2'(\phi) = \ket{\Phi(\phi)} \bra{\Phi(\phi)}=R_z(\phi) \ket{\Phi^+} \bra{\Phi^+} R_z^\dag(\phi)$, would yield the required measurements to reconstruct the subspace witness. Indeed, recall that to measure $\langle W_s \rangle$ from   the subspace-maximized Bell state fidelity $\text{max}_\phi[ F_\psi(\phi)]$ requires multiple fidelity measurements in order to learn the magnitude of the coherence, $|\rho_{k\bar{k}}|$. 
Similarly, if only one of the two qubits is observable, $M=\sigma_1^z$, one could use the correlator $\langle\sigma_1^\alpha\sigma_2^\alpha\rangle$ and their $R_z(\phi)$ rotations to extract $W_s$.

First, we note that $\sigma_1^z\sigma_2^z$ is invariant under $R_z$, and indeed it yields the population $P=(1+\langle \sigma_1^z\sigma_2^z\rangle)/4$, which is independent of $\phi$. To extract $C(\phi)$ here we used the same HHCP-based disentangling gate $U$ that generated the entanglement, resulting in the following signal $S=\langle M' \rangle$ up to a constant:
\begin{equation} \label{eq:MHH}
\begin{split}
S(\phi)  & =  \text{tr} (\rho U(\phi) M U^\dag(\phi)) \\
& = \frac{\langle \sigma_ 1^x\sigma_2^x - \sigma_1^y \sigma_2^y \rangle}{2} \cos(\phi) -  \frac{\langle \sigma_1^x \sigma_2^y +  \sigma_1^x \sigma_2^y \rangle}{2} \sin(\phi)\\
& = 2 \text{Re}(\rho_{14}) \cos(\phi) -  2 \text{Im}(\rho_{14}) \sin(\phi) \\
& = 2 C(\phi).
\end{split}
\end{equation}
Here $\rho_{14} = \langle00|\rho|11\rangle$ is the coherence of interest, and the (undesired) constant offset  under HHCP is given by $\langle \sigma_1^z-\sigma_2^z\rangle/2$, thus yielding a total of 4 measurement required to determine $W_s$ (while only three would be necessary if a joint projective measurement were available, see Appendix \ref{sec:appendix} for experimental details and signal derivation.)

\subsection{Experimental results}

We now discuss the measurement results when attempting to create the Bell state $\ket{\Phi^+}$, which results in the generation of the state $\rho$. 

First, we measure the state witness $\langle W_\Phi(\phi=0) \rangle$ which requires the measurement of the Bell state fidelity $\langle \Phi^+ |\rho| \Phi^+ \rangle= \frac14(1+\langle\sigma_1^z \sigma_2^z \rangle +\langle \sigma_1^x \sigma_2^x \rangle -\langle\sigma_1^y \sigma_2^y \rangle)$. 
From the 3 measurements shown in Fig.~\ref{fig:Figure2WitnessMeasurements}(a), we obtain $F_{\Phi^+}=1/4 (1 + 0.497 + 0.2142 + 0.5857)=0.57421(4)>1/2$. From this fidelity we have  $\langle W_\Phi \rangle = -0.07$ which successfully detects entanglement. 

Still, this measurement might underestimate the amount of entanglement, due to coherent, local errors. We thus  measure the subspace witness $\langle W_s \rangle$, to extract the  coherence $|\rho_{k\bar{k}} |  = |\langle00|\rho|11\rangle|$ (see Fig.~\ref{fig:Figure2WitnessMeasurements}(b) and Appendix \ref{sec:appendix} for experimental details and signal derivation.) 
We obtain a maximum fidelity of $\langle \Phi|\rho| \Phi \rangle = 0.6827(4)$, corresponding to $W_s=-0.1827(4)$, having optimized over the $|\Phi\rangle$ subspace. This indicates that we have indeed coherent errors affecting our state preparation and measurement process.

The subspace witness can also help determining the entanglement coherence time. 
The (entangled) prepared state $\rho(\tau\!=\!0)$ will evolve under the environment influence and the natural (or engineered) Hamiltonian. By measuring the subspace witness after a time $\tau$, one can then detect whether entanglement is still present, at the net of local, unitary evolution. By setting $\text{max}_\phi[F_\psi(\tau^*)] = \alpha$ we can further define a threshold time $\tau^*$ after which entanglement in $\rho(\tau\geq\tau^*)$ is no longer witnessed. 

To simplify the measurement of the subspace witness, we can apply the phase $\phi$ rotation at each time point measured, such that $\phi = \nu \tau$. Then, from the decay of the oscillations in $C$ one can directly extract  $W_s$ at $\tau=0$ and the characteristic time $\tau^*$. We note that in general both coherences $C$ and population $P$ will decay for open quantum systems, but only extracting the coherences at various $\phi$ is needed to reconstruct $W_s$. In our experiments (see Fig.~\ref{fig:Figure2WitnessMeasurements}.(c)), we thus measure the phase-modulated decay of the coherence. As the main decoherence source is dephasing, which leaves populations intact, we simply  verify that $P(\tau)$ is constant. 

%2019-09-03: In our experiments (see Fig.~\ref{fig:Figure2WitnessMeasurements}.c), we only measured the decay of the coherence, as the main decoherence source is dephasing which leaves populations intact. 
%
%todo: removed bound: However, for special states, one can bound the fidelity by its coherence: more specifically, given that general $\rho$ satisfies $ \rho_{jj} + \rho_{kk} \geq 2\sqrt{\rho_{jj}\rho_{kk}} \geq 2|\rho_{jk}|$ by Cauchy-Schwartz inequality, this yields $F_\psi = P+C \geq 2 C$ for Bell or GHZ states. This allows estimation of a lower-bound (i.e., more conservative) timescale $\tau^*_\text{LB} < \tau^*$ by solving for $2 C(\tau^*_\text{LB}) = \alpha$ where $\alpha=1/2$ for Bell or GHZ states. 
%
In experiments, we compare $P(\tau)\!=\!(1+\langle \sigma_1^z \sigma_2^z (\tau) \rangle)/4$ at $\tau=0\,\mu$s and $\tau=40\mu$s and, as expected, observe no decay in $\langle \sigma_1^z \sigma_2^z (\tau) \rangle$.
We then study the entanglement decay under a spin echo~\cite{Hahn50, Cooper18} of varying duration $\tau$.
The prepared entangled state $\rho$ evolves under $\mathcal{H}_\text{int}$ (which would not affect the ideal state $\ket{\Phi}$) and decoherence. We then apply  the disentangling gate (measurement) modulated at $\phi = \nu \tau$ with $\nu=15$ kHz.  A simple Gaussian decay fit yields a characteristic decoherence time $T_2=31(3)\mu s$ of the double-quantum coherence $|\langle00|\rho|11\rangle|$, such that taking a constant $P=0.374(1)$, we obtain the time $\tau^* =33(3)\mu s$ until which entanglement can be detected.

%todo: check + do propagation of error + change figure caption.

While we have shown that we can create entanglement in our system, the state fidelity is not optimal. To improve the quality of entanglement in our hybrid system and investigate the source of non-ideality, we probe the entanglement as a function of repetitive state initialization steps~\cite{fig:Fig3}. In this way, we can distinguish between initialization $I$ and control errors in $U$. 
 We repeat the HHCP plus laser polarization block   $N=\{1,3,5\}$ times to create $\rho_0(N)$,  then prepare the entangled state $\rho(N)$ and measure the witness $\langle W_s(N) \rangle$ with fixed control operations. With increasing $N$ we observe an increase in the overall fidelity $F_s(N)$, due to an increase in both the population difference (inferred from $P(N)$) as well as the double-quantum coherence $C(N)$. We note that explicitly verifying an increase in coherence is of practical importance, because specific applications, e.g., classical field sensing with GHZ states, depend strictly on the amount of coherence (not necessarily fidelity) of the entangled state.

\begin{figure}
  	\includegraphics[width=\columnwidth]{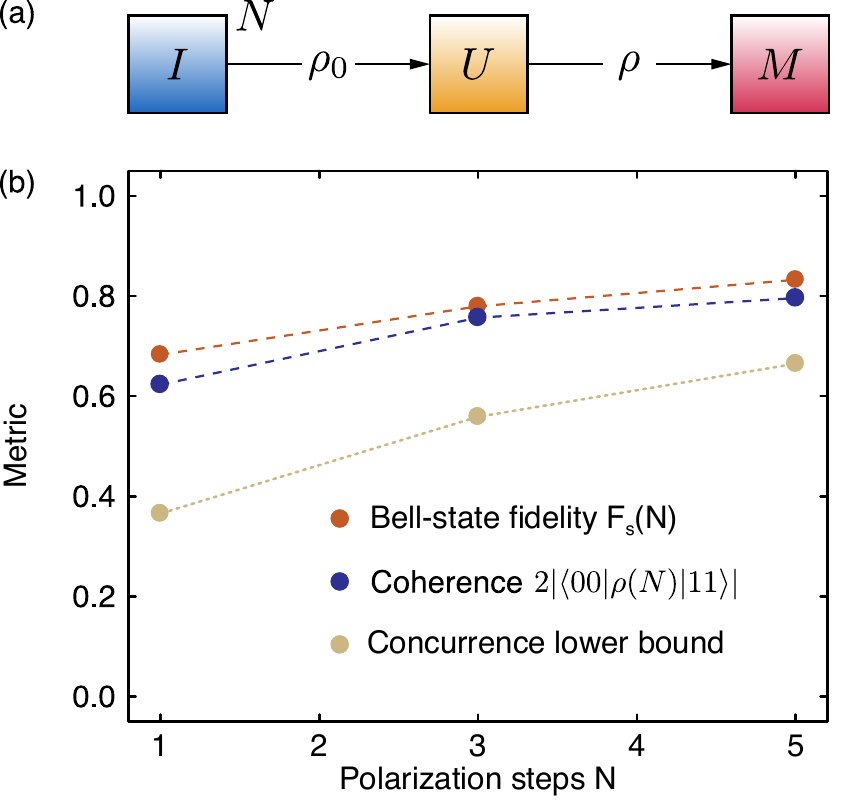}
 	\caption{\textbf{Improved entanglement detection allows improved bound to entanglement}: 
	(a) Given imperfect state initialization step that prepares $\rho_0$ with subunit purity, it is possible to improve purity by $N$ repetitive initialization steps. 
	(b) We plot as a function of $N$ the following results of the subspace witness: namely, the Bell state fidelity $F_s(N) =P(N)+|\langle 00| \rho (N) | 11\rangle|$, the coherence $2|\langle 00| \rho (N) | 11\rangle|=2 \max_\phi[C(\phi)] \leq F_s(N)$, and the resulting lower bound to concurrence. With $N>1$ we observe the expected improvement in purity, from $P(N)=(1+\langle \sigma_1^z \sigma_2^z (N) \rangle)/4$, that leads to improved $F_s(N)$. We also verify the increase in double-quantum coherence generated $|\langle 00| \rho (N) | 11\rangle|$, which is of practical importance given that specific applications such as entanglement-enhanced sensing with GHZ states benefit directly from larger quantum coherence and not directly the fidelity itself. In addition, we note that the subspace witness $\langle W_s\rangle$ improves bound to entanglement (via concurrence) over the typical `state' witness $\langle W_\psi \rangle$ due to improved Bell state fidelity. The errorbars are smaller than the dots for all plots. The applicability of improved bound for specific GME states by $\langle W_s\rangle$ is discussed in the text.}
	\label{fig:Figure3LowerBoundEntanglement}
\end{figure}

Finally, we note that the subspace witness provides stricter bound on the amount of  entanglement generated. While this might seem intuitive,
we remark  that  a `general' way to relate entanglement detection to  existing quantifiers is not known. 
However, for the two-qubit case, it has been shown~\cite{Bennett96} that one can relate Bell fidelities $F_\psi$  to the entanglement of formation, and thereby to any other related metrics such as the well-known concurrence. More specifically, the lower-bound to the two-qubit concurrence is $C_2(\rho) \geq \max (0, -2W_\psi)$, with $C_2=1$ for maximally entangled states. 
This relation makes it clear that to obtain an entanglement measure one should optimize over all state witnesses. While the subspace witness only optimizes over a restricted set of states, it still provides a stricter bound than the state witness, $C_2\geq -2W_s\geq -2W_\psi$.

\section{Extension to specific genuine multipartite entangled (GME) states}
\label{sec:GME}
We wish to extend the notion of subspace witness to those multipartite entangled states that allow entanglement detection by witnesses $W_\psi$. 
%We will see that the method not only improves entanglement detection as expected while remaining efficient with respect to QST, but also allows lower-bound quantification by the GME concurrence~\cite{GMEconcurrence} which in contrast would not be allowed by a single state witness measurement.
%
To this end, we first parameterize a multipartite entangled state $|\psi(\vec{\phi})\rangle$ in the computational basis $|k\rangle$ as:
\begin{equation} \label{eq:targetEntangledState}
\begin{split}
|\psi (\vec{\phi}) \rangle & = \sum_{k=0}^{d-1} a_k e^{-i \phi_k} |\vec{k}\rangle,
\end{split}
\end{equation}
where $a_k$ is the probability amplitude, $\phi_k = \vec{k}\cdot\vec{\phi}$ is the $|\vec{k}\rangle$-dependent phase given a preset $n$-length vector of phases $\vec{\phi} = \{\phi_j\}^n$ for every $j$-th qubit, and $d$ is the dimension of the subspace spanned by the set $\{ \ket{k}\}$ specifying $\ket{\psi}$. While such expression could describe any pure state of $n$ qubits, we are interested in NPT entangled states $|\psi\rangle$ for which state witnesses $W_\psi$ are valid, such as GHZ, W, or Dicke states. For instance, for a general $n$-qubit GHZ state $\ket{\psi_k}=(\ket{k} + e^{-i \phi} \ket{\bar{k}})/\sqrt{2}$, we have $a_k = 1/\sqrt{2}$ and the subspace of interest is spanned by $\{ \ket{k}, \ket{\bar{k}}\}$ of dimension $d=2$. For a W state $\ket{W(n)}=\sum_{k=1}^n e^{-i \phi_k} \ket{k}/\sqrt{n}$, we have $a_k = 1/\sqrt{n}$ with the subspace of dimension $d=n$ spanned by states $\{\ket{k}\}$ in the one-excitation manifold.

Given this parameterization, the fidelity $F_\psi$ again reduces to a simple expression $F_\psi(\vec{\phi}) = P + C(\vec{\phi})$ with $\vec\phi=\{\phi_k\}$ for $k=1,\dots,d$. Here,
$P=\sum_{k=0}^{d-1} a_k^2 \rho_{kk}$ and $C(\vec{\phi})$ is just a sum similar to  Eq.~\ref{eq:BellCoherence} extended to all many-body coherences $\rho_{jk}$ of interest:
\begin{equation} \label{eq:explicitSubspaceWitness}
\begin{split}
C(\vec{\phi}) & = 2 \sum_{j=0}^{d-1} \sum_{k>j}^{d} a_j a_k  \vec{c}_{jk} \cdot \vec{o}_{kj}, 
\end{split}
\end{equation}
where $\vec{c}_{jk} = (\text{Re} (  \rho_{jk} ), \text{Im} ( \rho_{jk} )) $ and $\vec{o}_{kj}= (\cos \phi_{kj}, \sin \phi_{kj} )$ with $\phi_{kj} = \phi_k - \phi_j$. 

As in the $n=2$-qubit case, to extract the `subspace' witness $\langle W_s\rangle$, we must identify all $d(d-1)/2$ many-body coherences $\rho_{jk}$ by again solving the set of linear equations given by multiple measurements $F_\psi(\vec{\phi})$, containing a total of $d(d-1) +1$ unknowns. 

Therefore the subspace dimension $d$, and its scaling with the number of qubits $n$, sets the limit  of which entangled states we can tackle, given that we want to be efficient with respect to QST. 

For GHZ states, $d=2$ is constant and independent of $n$: in other words, for any $n$-qubit  quantum processor, one can extract the subspace-maximized GHZ witness with just $3$ measurements, a very efficient protocol.  Indeed, experimentally such subspace-optimized fidelity has been observed with $\sim20$ qubits in superconducting and neutral atom systems~\cite{Wei19, Omran19, Song19}. We also note that using a 10-qubit register in diamond up to a 7-qubit GHZ state was witnessed with state fidelity $0.589(5)$~\cite{Bradley19} which could be further optimized with such subspace witness measurement.

For W states, $d=n$: since this is still polynomial in $n$, all subspace-optimized $W_s$ witnesses will be efficient with respect to QST. 
In contrast,  for Dicke states, $d= {{n}\choose{k} } = {{n}\choose{n-k} }$ with $k$ excitations: therefore, only for very lowly- (highly-) excited Dicke states will $\langle W_s \rangle$ prove efficient with respect to QST.

While for concreteness we give examples with these well known GME states, they also highlight the motivation behind improving entanglement detection via subspace witnesses (as alluded to above with the two distinct Bell subspaces): namely that the amount of useful entanglement for the specific QIP task, i.e., considering entanglement as a resource, often depends on the total magnitude of the quantum coherence within that subspace, independent of relative phases that could result from unknown local SPAM errors. This idea is reflected in entanglement quantifiers, which is independent of local unitary formations, but more concretely seen in well-known QIP tasks of interest. For instance, in entanglement-enhanced sensing of classical fields, it can be shown that the sensitivity depends on the magnitude of the many-body coherence in the GHZ subspace; or for improved quantum memories via decoherence-free subspace, e.g., with W-like states, the lifetime of quantum information depends on the magnitude of the coherences. Thus, the subspace witness, while alleviating the problem of local unitary control errors present during entanglement detection measurements, better conveys the notion of entanglement as a resource, almost as an intermediate between entanglement detectors and quantifiers. 
%Before concluding, we note that the subspace witness could be seen to more accurately convey the notion of entanglement as a resource. This is not only shown through enabling lower bound quantification of entanglement (given population measurements), but also from specific examples of QIP tasks for which the target entangled states are useful. For instance, for entanglement-enhanced sensing, it can be shown that the sensitivity depends on the magnitude of the coherence in the GHZ subspace; and for improving quantum memories via decoherence-free subspace, e.g., with W-like states, the lifetime of quantum information depends on the (initial) magnitude of the coherences. Therefore, the subspace witness, while addressing the problem of local unitary control errors present during entanglement detection measurements, could more faithfully report the amount of useful entanglement.

In fact, this link could be made more concrete by noting that the subspace witness also facilitates quantification of lower bound of entanglement, with additional measurements of population terms.
%Finally, we remark that the subspace witnesses emphasize the importance of measuring the magnitude of coherences in characterizing entanglement. This is  related to the role of coherences not only in some quantum applications such as entanglement-enhanced sensing, but also  in bounding the amount of entanglement for some GME states. 
%could facilitate the lower-bound quantification of entanglement for the specific GME states. 
More specifically, the lower bound to an entanglement measure called the GME concurrence $C_\text{GME}$~\cite{Ma11}, related to the separability criteria~\cite{Guhne10}, can be estimated efficiently from experiments as it requires the knowledge of only specific matrix elements of $\rho$. 
Both the lower bound to $C_\text{GME}$ and separability criteria take the form of a difference between the many-body coherences $\rho_{jk}$ within the subspace of interest and appropriate population terms outside the subspace. 
Similar to entanglement witnesses, these quantities change sign for separable states, as the difference between coherences and population terms changes sign.

%%%now turn to lower-bound quantification of entanglement, which is realized by such subspace witness measurement. While this task is very difficult in general for many existing quantifier, past works on separability/GME criteria~\cite{GMEcriteria1, GMEcriteria2} have led to an experimentally implementable lower-bound to the measure called GME concurrence~\cite{GMEconcurrence}, which only requires the knowledge of specific many-body coherences of interest and relevant population terms of $\rho$. 

For instance, for $n=3$, the lower bound to GME concurrence is given by:
\begin{equation} \label{eq:3qubitGMEConc}
\begin{split}
C_\text{GME}(n=3) \geq \rho_{18}-\sum_{k, \bar{k} \neq 1,8} \sqrt{\rho_{kk} \rho_{\bar{k}\bar{k}}},\\
\end{split}
\end{equation}
where the first term is $\rho_{18}=\langle 000| \rho| 111 \rangle$, and the second term is the sum of populations outside the GHZ subspace. 

We note that the subspace witness $\langle W_s \rangle$ alone is insufficient in providing the lower bound to GME concurrence: $\langle W_s \rangle$ can only provide the first term, as it identifies the true (maximum) coherences of interest. Thus the second term of populations must be identified from additional measurements, but for systems with individual qubit readout, a single measurement setting (every qubit along $z$) suffices to identify all the population terms.

%%%For more established $n=3$, there exist other computable measures such as the monotone 3-tangle $\tau_3$ for arbitrary states that may not require exact identification of all many-body coherences~\cite{3tangle}:

%%%\begin{equation} \label{eq:3tangleApprox}
%%%\begin{split}
%%%\tau_3^\text{approx}(\rho) & = \max(0,  \pm\frac{8}{7} C(\phi) \pm\frac{20}{7}P - 3 ) \\
%%%\end{split}
%%%\end{equation}

%%%However, for larger $n$, there are as of yet not many easily estimatable quantifiers, and thus the identification of every many-body coherence of interest as enabled by $\langle W_s \rangle$ allows a step towards this task. Finally, we note that the relevant population terms will need to be measured in addition, but for systems with individual qubit readout can obtain all the population terms in one measurement setting (every qubit along $z$).

\section{Conclusion}
Typical entanglement witnesses based on fidelity measurements to target entangled states $|\psi\rangle$---which we call `state' witnesses $\langle W_\psi \rangle$---while being efficient detectors of entanglement, can underestimate actual entanglement present in $\rho$ due to local, unitary errors in state preparation and measurement (SPAM). Therefore, we develop the idea of `subspace' witness $\langle W_s\rangle$, based on two-qubit systems, which strictly observes a larger amount of witness violation than does $\langle W_\psi \rangle$ at the cost of additional measurements, while being efficient with respect to state tomography. Conceptually, the appropriate subspace is chosen for relevance to particular quantum information tasks, and because the subspace witness yields a value optimized over local unitaries within the subspace, it could be viewed as an intermediate metric between an entanglement measure (invariant under local unitaries) and a typical witness (dependent on local unitaries). Experimentally, using a two-qubit solid-state system composed of electronic spins, we observe a significantly improved detection by the subspace witness, motivating the use even for small quantum systems which may have non-negligible local unitary errors. Finally, we extend the subspace witness to improve detection for specific genuine multipartite entangled (GME) states, which may aid near-term NISQ devices to better characterize their performance for applications involving specific entangled states of interest. Because the subspace witness essentially identifies the true (i.e., maximum) many-body coherences of interest, it not only guarantees improved entanglement witness detection but also aids in improving lower bounds to entanglement via experimentally-friendly metrics such as GME concurrence or the separability criteria, which require the knowledge of only specific matrix elements of $\rho$.

\section{Acknowledgements}
This work was in part supported by NSF Grants No. PHY1415345 and No. EECS1702716.

\appendix

\section{Experimental details and signal derivation}
\label{sec:appendix}

\subsection{State witness $\langle W_\psi \rangle$ measurement}
%\textbf{keep notation consistent: if you used $\sigma_\alpha$ above, use them everywhere.}

Our experimental system is a hybrid $n=2$-qubit system in which only the first qubit can be directly observed with the (bare) operator $M = \sigma_1^z \otimes \openone$, with Pauli operators $\sigma_j^{\{x,y,z\}}$ on the $j$-th qubit. Thus to measure any two-body correlator $\langle \sigma_1^\alpha \sigma_2^\beta \rangle$ (e.g., to reconstruct Bell state fidelities), one can evolve $M$ under a two-qubit gate. Given that in our system the qubits interact by $\mathcal{H}_\text{int}=d \sigma_1^z \sigma_2^z /2$, a simple experimental sequence involving only single-qubit $\pi/2$-rotations and free evolution under $\mathcal{H}_\text{int}$ yields any desired two-qubit correlator. In other words, first defining a single-qubit rotation on $j$-th qubit along axis $\phi$ by $\theta$ as $R_{\phi,j}(\theta)=e^{-i \theta/2 (\sigma_j^x \cos(\phi) + \sigma_j^y \sin(\phi))}$, a simple pulse $U(\phi) = R_{\phi}(\pi/2)\cdot e^{-i \mathcal{H}_\text{int} t} \cdot R_{\pi/2}(\pi/2)$ will suffice. To mitigate dephasing during free evolution however, one can also insert global $\pi$-pulse(s) on all qubits during the free evolution so as to decouple from the environment bath of spins. Therefore, we insert one $\pi$-pulse in the middle of the free evolution for both NV and X spin, resulting in $U(\phi) = R_{\phi}(\pi/2)\cdot e^{-i \mathcal{H}_\text{int} t/2} \sigma_1^y\sigma_2^y e^{-i \mathcal{H}_\text{int} t/2} \cdot R_{\pi/2}(\pi/2)$. Therefore the effective measurement operator $M'$ is
\begin{equation} \label{eq:measOpZZ}
\begin{split}
M'  	= & U(\phi)MU(\phi)^\dag,\\
	= &  \cos(\phi)[\sigma_1^y \cos(dt) + \sigma_1^z \sigma_2^z\sin(dt)] + \\
	&  \sin(\phi)[\sigma_1^z\cos(dt) - \sigma_1^y\sigma_2^z\sin(dt)].\\
\end{split}
\end{equation}

Thus overlapping $\rho$ with $M'$ at the optimal time $t=(2\pi)/(4d)$ with $\phi=0$ gives the desired signal $S$:
\begin{equation} \label{eq:measSigZZ}
\begin{split}
S & = \text{tr}(\rho M')\\
& =  \langle \sigma_1^z\sigma_2^z\rangle.\\
\end{split}
\end{equation}

\subsection{Subspace witness $\langle W_s \rangle$ measurement}

Measuring the fidelity $F_\psi(\phi)=P+C(\phi)$ similarly involves evolving the bare measurement operator $M$ under a combination of single-qubit rotations and two-qubit gates. In the experimental work shown, we utilize HHCP to realize the two-qubit gate to both generate and detect entanglement by the subspace witness. More specifically, the evolution under HHCP can be described by the Hamiltonian $\mathcal{H}_\text{HH} = d ( \sigma_ 1^x\sigma_2^x \pm \sigma_1^y \sigma_2^y)/4$, in which the sign determines in which subspace (either $-$ for $\ket{\Phi}$ or $+$ for $\ket{\Psi}$) the evolution will happen. For instance, choosing the $\ket{\Phi}$ subspace, the bare measurement operator under the pulse sequence $U(\phi) = R_z(\phi) e^{-i \mathcal{H}_\text{HH} t} R_z^\dag (\phi)$ evolves to the desired two-body operators:
\begin{equation} \label{eq:measOpHH}
\begin{split}
2M'(\phi) = & 2 U(\phi) M U^\dag(\phi)\\
 = & + (\sigma_ 1^z-\sigma_2^z) \\
& + (\sigma_ 1^z+\sigma_2^z)\cos(dt) \\
& +  (\sigma_ 1^x\sigma_2^x - \sigma_1^y \sigma_2^y)\cos(\phi)\sin(dt) \\
& - (\sigma_1^x \sigma_2^y +  \sigma_1^x \sigma_2^y)\sin(\phi) \sin(dt).
\end{split}
\end{equation}

%& = \frac{\langle \sigma_ 1^x\sigma_2^x - \sigma_1^y \sigma_2^y \rangle}{2} \cos(\phi) -  \frac{\langle \sigma_1^x \sigma_2^y +  \sigma_1^x \sigma_2^y \rangle}{2} \sin(\phi)\\
Taking the overlap of $\rho$ with $M'(\phi)$ at the optimal time $t=(2\pi)/(4d)$ gives the signal $S$ carrying the desired information on the two-body coherence:
%\textbf{do not leave a line between text and equation. Equations are part of the text, you would not leave a blank line between two parts of a sentence. Also, put commas (or period) as required at the end of the equations.}
\begin{equation} \label{eq:measSigHH}
\begin{split}
2 S(\phi)  = & 2 \text{tr}(\rho M'(\phi))\\
= & + \langle \sigma_ 1^z-\sigma_2^z\rangle \\
& +  \langle \sigma_ 1^x\sigma_2^x - \sigma_1^y \sigma_2^y \rangle \cos(\phi) \\
& - \langle \sigma_1^x \sigma_2^y +  \sigma_1^x \sigma_2^y\rangle \sin(\phi)\\
= & +  \langle \sigma_ 1^z-\sigma_2^z\rangle \\
& + 4\text{Re}(\langle00|\rho|11\rangle) \cos(\phi) \\
& + 4\text{Im}(\langle00|\rho|11\rangle)\rangle \sin(\phi)\\
= & + \langle \sigma_ 1^z-\sigma_2^z\rangle + 4 C(\phi),
\end{split}
\end{equation}
where the first term makes up the (undesired) constant.

\section{Example for Bell States}
\label{sec:appendixB}

Here we give an example for the two qubit case (for which an analytical expression of entanglement measure $C_2$ can be obtained) that shows that the subspace entanglement witness $\langle W_s \rangle$ detects a larger share of entangled states than does a typical state witness $\langle W_\psi\rangle$, also with a larger violation.

Consider a generic state in the subspace spanned by one pair of Bell State (e.g., $\ket{\Phi^\pm}$). The state can be written as
\[\rho_\Phi=\frac{\openone_\Phi}2+\frac\epsilon2(\sin\phi_0\cos\theta\,\Phi_x+\sin\phi_0\sin\theta\,\Phi_y+\cos\phi_0\,\Phi_z),\]
where $\Phi_\alpha$ are Pauli matrices in the (sub)space, e.g., $\Phi_x=\ket{\Phi^+}\!\!\bra{\Phi^-}+\ket{\Phi^-}\!\!\bra{\Phi^+}$ and $\openone_\Phi$ is the identity in the subspace. $\rho_\Phi(\epsilon, \theta, \phi_0)$ is uniquely defined in the following range: $\epsilon\geq0$, $\theta \in [0, 2\pi)$, and $\phi_0\in[0,\pi]$, where $\epsilon=0$ indicates a classical mixture. To gain a bit of insight into this generic state in the chosen subspace, we note that the double-quantum coherence $C$ (necessarily nonzero for entanglement) is given by
\begin{equation} 
\begin{split}
C = \langle 00|\rho_\Phi(\epsilon, \theta, \phi_0) | 11\rangle = &  (\epsilon/2) (\cos\phi_0 + i \sin\phi_0\sin\theta).\\
\end{split}
\end{equation}
This shows that maximum entanglement at given $\epsilon$---at $|C|=\epsilon/2$---occurs either when $\sin^2\phi_0=0$ or $\sin^2\theta=1$. The former indicates a state with fully real coherence $C=\epsilon/2$, and the latter case refers to a more general complex coherence $C=e^{i\phi_0}\epsilon/2$.

Now we calculate the concurrence $C_2$ for this state which reveals that the state $\rho_\Phi(\epsilon>0, \theta, \phi_0)$ is in general entangled (except at a special point at $\theta=0, \phi_0=\pi/2$). More specifically, given $C_2 = \sqrt{a + b|\epsilon|} - \sqrt{a - b|\epsilon|}$, where
\begin{equation} 
\begin{split}
a 	= & (1 - \epsilon^2\sin^2\phi_0(1+\cos2\theta))/4;\\
b	= & \sqrt{b_1 b_2}/2,\\
b_1 	= & 2 - \sin^2\phi_0(1+\cos2\theta)\geq0,\\
b_2 	= & 2 - \sin^2\phi_0(1+\cos2\theta)\epsilon^2\geq0.\\
\end{split}
\end{equation}
where the equality for $b_{1,2}=0$ is achieved at the mentioned special point at $\epsilon=1$.
Since $C_2 > 0$, which simplifies to $2b |\epsilon| >  0$, indicates entanglement, we see in general $\rho_\Phi(\epsilon>0, \theta, \phi_0)$ is entangled due to positivity of $b$. Therefore, an ideal entanglement witness should detect all of $\rho_\Phi(\epsilon>0, \theta, \phi_0)$ (except at the special point).

Now let us first examine the state witness $\langle W_\psi \rangle$ for the `range' of states it can detect as well as its violation. Here the target state is either $|\psi\rangle=|\Phi^\pm\rangle$; we choose $|\Phi^+\rangle$ as it makes no difference in the detectable range or the degree of violation. More specifically, we see that
\begin{equation} \begin{split}
\langle W_\psi \rangle 	= & \frac12 -  \langle \Phi^+ | \rho_\Phi | \Phi^+ \rangle \\
					= & - \frac{\epsilon}{2} \cos\phi_0\\
 					= & - \text{Re}({\langle 00 | \rho_\Phi(\epsilon, \theta, \phi_0) | 11\rangle }). \\
%=& - \epsilon \cos\phi_0/2. \\
\end{split}
\end{equation}
Therefore, assuming $\epsilon \neq0$, the `range' of detectable states for $\langle W_\psi \rangle$ is $\phi_0\in[0,\pi/2)$, while all of $\phi_0\in[0,\pi]$ are entangled as seen from concurrence. Furthermore, the state witness, being oblivious to $\theta$, will underestimate or completely miss all the entanglement in the imaginary part of the coherence. 

Finally, we show that the subspace witness $\langle W_s \rangle$ captures a larger share of entangled states, namely all the entangled states in the subspace, also with a larger violation. The subspace witness is 
\begin{equation} \begin{split}
\langle W_s \rangle 	 = &  \frac12 - \max_\phi \langle \Phi (\phi) | \rho_\Phi | \Phi(\phi) \rangle \\
=&-\frac \epsilon2 \max_\phi\left\{\sin\theta \sin\phi_0  \sin\phi+\cos \phi_0  \cos\phi\right\} \\
=&-\frac \epsilon2  (\sin^2\theta \sin^2\phi_0 + \cos^2 \phi_0)^{1/2} \max_\phi\left\{   \cos(\phi-\theta')\right\}\\
=&- |\langle 00 | \rho_\Phi(\epsilon, \theta, \phi_0) | 11\rangle | \max_\phi\left\{   \cos(\phi-\theta')\right\}\\
=&- |\langle 00 | \rho_\Phi(\epsilon, \theta, \phi_0) | 11\rangle |.
\end{split}
\end{equation}
%%We note that for any state with $\epsilon>0$, and  $\theta\neq0,\pi$ or $\phi\neq\pi/2+k\pi$, we can always find a value for $\phi$ such that the witness is negative, thus detecting entanglement for all entangled states in the subspace. 
In other words, the subspace witness detects with maximum violation $|C|$ all of $\rho_\Phi(\epsilon>0, \theta, \phi_0)$ except at the special (unentangled) point mentioned above. 
We make one note regarding the analytical maximization in the third equality is realized at a single $\phi$ since there is only a single coherence $\langle 00 |\rho|11\rangle$ to overlap with. Of course, experimentally one knows not the optimal $\phi$ a priori, so as discussed in main text, such dimension $d=2$ entangled states require 3 measurements to measure the subspace witness. An explicit example to measure the subspace witness with multiple coherence terms is discussed in the next section.

\section{Example $\langle W_s \rangle$ measurement scheme for target $|W(n=4)\rangle$}
\label{sec:appendixC}

Recall that a single fidelity measurement yields $\langle \psi ({\phi}) |\rho |\psi ({\phi}) \rangle = P + C ({\phi})$, where  
\begin{equation} \label{eq:Cphi}
\begin{split}
C(\vec{\phi}) & = 2 \sum_{j=0}^{d-1} \sum_{k>j}^{d} a_j a_k  \vec{c}_{jk} \cdot \vec{o}_{kj}\\
 & = 2 \sum_{j=0}^{d-1} \sum_{k>j}^{d} a_j a_k  (\text{Re}(\rho_{jk})\cos(\phi_{kj}) + \text{Im}(\rho_{jk})\sin(\phi_{kj})),\\
\end{split}
\end{equation}
with $\phi_{kj} = \phi_k - \phi_j$, where $\phi_k = \vec{k}\cdot\vec{\phi}$ is the state $|\vec{k}\rangle$-dependent phase given a preset $n$-length vector $\vec{\phi} = \{\theta_m\}^n$, with the phase $\theta_m$ on the $m$-th qubit in general. Here, assuming control over the $n$-length vector $\vec{\phi}$ assumes universal control over all qubits such that individual single-qubit control can imprint an arbitrary phase $\theta_m$ along z. 

Control over individual qubit phases $\theta_m$ allows a simple method to reconstruct the subspace-optimized fidelity and thus improved entanglement detection. For instance, consider the general definition of the W-entangled state $\ket{W(n)}=\sum_{k}^n e^{-i \phi_k} \ket{k}/\sqrt{n}$. Given single-qubit phase control $\vec{\phi} = \{\theta_m\}^n$, the W-state can be written in a more experimentally-friendly manner as $\ket{W(n)}=\sum_{m=1}^n e^{-i \theta_m} \ket{m}/\sqrt{n}$, where $m$ indicates excitation on the $m$-th qubit. This `practical' definition allows simple parameterization of the fidelity $F(\vec{\phi})$ measurements so as to carefully (with minimal measurements) reconstruct $\max_{\vec{\phi}} [F(\vec{\phi})]$.

Given multiple ways to reconstruct $\max_{\vec{\phi}} [F(\vec{\phi})]= P + \max_{\vec{\phi}}[C(\vec{\phi})]$, here we discuss one simple approach to reconstruct $\max_{\vec{\phi}}[C(\vec{\phi})]$. First, we note that a judicious choice of $\vec{\phi}$, as seen from Eq.~\ref{eq:Cphi}, decouples equations containing either only the real or only the imaginary parts of coherences $\rho_{jk}$. 
More specifically, $C(\vec{\phi})$ contains either only $\text{Re}(\rho_{jk})$ or $\text{Im}(\rho_{jk})$ if one measures either $\phi_k=\{0, \pi\}$ or $\phi_k=\{\pm \frac{\pi}{2} \}$ respectively. 
Notice this (arbitrary) restriction to a binary set will accordingly reduce the maximum number of unique fidelity measurements (equations) available to solve for $|\rho_{jk}|$. 
More specifically, given $d$ $|k\rangle$ vectors describing the target $|\psi\rangle$, there are now $(d-1)$ relative phases $e^{-i \phi_k}$; therefore the binary set of inputs allows at most $2^{d-1}$ unique fidelity measurements (equations). 
Since the number of unknown parameters needed to identify  $|\psi\rangle$ is $d(d-1)+1$, one must make sure the available equations (in this case $2^{d-1}$) outnumber $d(d-1)+1$. 
The binary restriction of inputs allows this for GHZ, W, and a subset of lowly (highly) excited Dicke states. 
Instead, the subspace dimension $d$ of intermediately-excited Dicke states increases non-polynomially with qubit number $n$, such that $d(d-1)+1 > 2^{d-1}$. In this case,  the number of measurements required for  entanglement witnessing tends towards that required for state tomography, thus defeating the very purpose of entanglement witnesses.

%Then, by permuting the possible input vector $\phi \equiv \phi_1\phi_2...\phi_{d-1}$---more specifically making $d(d-1)/2+1$ measurements out of $2^{d-1}$ possible choices, one can solve for every real (imaginary) part of the coherences to reconstruct the maximum coherence $|\rho_{jk}|$. 

Finally, we explicitly describe the subspace witness measurement scheme for a $n\!=\!4$ qubit W-entangled state: 
%$|\psi(\vec{\phi})\rangle=(|0000\rangle+e^{-i \phi_1}|0010\rangle+e^{-i \phi_2}|0100\rangle+e^{-i \phi_3}|1000\rangle)/2=(|0\rangle+e^{-i \theta_1}|1\rangle+e^{-i \theta_2}|2\rangle+e^{-i \theta_3}|3\rangle)/2$. %As discussed above, while $\phi_k$ can be realized in multiple ways, here for simplicity let $\phi_k
\begin{equation} \label{eq:4qubitWstate}
\begin{split}
|\psi(\vec{\phi})\rangle 	& = (|0001\rangle + e^{-i \phi_1} |0010\rangle + e^{-i \phi_2} |0100\rangle + e^{-i \phi_3} |1000\rangle)/2 \\
			 		& \equiv (|0\rangle + e^{-i \theta_1} |1\rangle + e^{-i \theta_2} |2\rangle + e^{-i \theta_3} |3\rangle)/2. \\
\end{split}
\end{equation}
The second equation shows how   for simplicity the phase $\phi_k$ can be realized by a phase $\theta_m$ on the $m$-th qubit, as discussed above.
%with normalization amplitude $a_k=n^{-1/2}$ (not shown). 
Thus each fidelity measurement is parameterized by $\vec{\phi} = \{\theta_1,\theta_2,\theta_3\}$, where we removed $\theta_0$ as it simply gives rise to a global phase.  Then, defining $R_{jk} \equiv 2 a_j a_k \text{Re}(\rho_{jk}) = \text{Re}(\rho_{jk}) /2$, each fidelity measurement yields $F_\psi(\vec{\phi}) = P + C(\vec{\phi})$ where
\begin{equation} \label{eq:4qubitWstateFidelity}
\begin{split}
C(\vec{\phi}) & = \sum_{j=0}^{d-1} \sum_{k>j}^{d} (-1)^{(\phi_k \oplus \phi_j)} R_{jk}.\\
\end{split}
\end{equation}
Thus to extract $\langle W_s \rangle$ we must first identify all $R_{jk}$. 

For convenience, let us define the sum and difference of two fidelity measurements as $y_\pm(\vec{\phi},\vec{\phi}') = F_\psi(\vec{\phi}) \pm F_\psi(\vec{\phi}')$. Then choosing the set of inputs $\vec{\phi}=\{001, 010, 100\}$ and $\vec{\phi}' =\vec{\phi} \oplus \vec{1} $ (i.e., the spin-flipped state), we identify 3 out of the 6 real parts $\{ R_{01}, R_{12}, R_{03} \}$ via the differences of fidelities:
\begin{equation} \label{eq:4qubitWstateFidelityDiffExamles}
\begin{split}
y_-(001, 110) 	& = 2 (+R_{01} + R_{02} - R_{03}) \\
y_-(010, 101) 	& = 2 (+R_{01} - R_{02} + R_{03}) \\
y_-(100, 011) 	& = 2 (-R_{01} + R_{02} + R_{03}). \\
\end{split}
\end{equation}

Then, by the sum of fidelities we almost identify the remaining 3 real parts $\{ R_{12}, R_{13}, R_{23} \}$:
\begin{equation} \label{eq:4qubitWstateFidelitySumExamles}
\begin{split}
y_+(001, 110) 	& = 2 (P + R_{12} - R_{13} - R_{23}) \\
y_+(010, 101) 	& = 2 (P - R_{12} + R_{13} - R_{23}) \\
y_+(100, 011 ) 	& = 2 (P - R_{12} - R_{13} + R_{23}). \\
\end{split}
\end{equation}
More specifically, we see that (due to unknown $P$) we need one more measurement, e.g. at $\vec{\phi}=\{0,0,0\}$, to solve for the remaining unknowns.

Therefore, a total of $d(d-1)/2 + 1=6+1=7$ measurements identifies all 6 real parts of $\rho_{jk}$ and P. In the same manner, $d(d-1)/2=6$ more measurements at $\phi_k=\{\pm\frac{\pi}{2} \}\equiv\{0,1\}$ will identify the imaginary parts of $\rho_{jk}$, such that one can reconstruct  $\langle W_s \rangle = \alpha - \max_{\vec{\phi}}[F_\psi(\vec{\phi})]=\alpha -  (P + 2 \sum_{j=0}^{d-1}\sum_{k>j}^d a_j a_k |\rho_{jk}|$).

\bibliographystyle{apsrev4-1}
\bibliography{subspaceWitnessBiblio}

\end{document}